\documentclass{article}
\usepackage{algorithm}
\usepackage[algo2e, ruled,vlined]{algorithm2e}
\usepackage{amsmath}
\usepackage{booktabs}
\usepackage{dirtytalk}
\usepackage{graphicx}
\usepackage{mlspconf}
\usepackage{multicol}
\usepackage{multirow}
\usepackage[dvipsnames]{xcolor}
\usepackage{setspace}
\usepackage{subfigure}
\usepackage{tabularx}
\usepackage{url}
\usepackage{soul}

\copyrightnotice{U.S.\ Government work not protected by U.S.\ copyright}

\copyrightnotice{979-8-3503-2411-2/23/\$31.00 {\copyright}2023 Crown}

\copyrightnotice{979-8-3503-2411-2/23/\$31.00 {\copyright}2023 European Union}

\copyrightnotice{979-8-3503-2411-2/23/\$31.00 {\copyright}2023 IEEE}

\toappear{2023 IEEE International Workshop on Machine Learning for Signal Processing, Sept.\ 17--20, 2023, Rome, Italy}

\title{An Enhanced System for the Detection and Active Cancellation \\ of Snoring Signals}
\name{%
   V. Bruschi$^{\star}$%
   \quad M. Cantarini$^{\star}$%
   \quad L. Serafini$^{\star}$%
   \quad S. Nobili$^{\dagger}$%
   \quad S. Cecchi$^{\star}$%
   \quad S. Squartini$^{\star}$\thanks{This work was supported by the financial program DM MiSE 5 Marzo 2018, project ``\emph{ChAALenge}''---F/180016/01-05/X43.}%
}
\address{%
   $^{\star}$ Department of Information Engineering - Università Politecnica delle Marche, Ancona, Italy \\%
   $^{\dagger}$ Leaff Engineering Srl, Ancona, Italy%
}

\begin{document}

\maketitle

\begin{abstract}
Snoring is a common disorder that affects people's social and marital lives. The annoyance caused by snoring can be partially solved with active noise control systems. In this context, the present work aims at introducing an enhanced system based on the use of a convolutional recurrent neural network for snoring activity detection and a delayless subband approach for active snoring cancellation.
Thanks to several experiments conducted using real snoring signals, this work shows that the active snoring cancellation system achieves better performance when the snoring activity detection stage is turned on, demonstrating the beneficial effect of a preliminary snoring detection stage in the perspective of snoring cancellation.
\end{abstract}


\section{Introduction}\label{sec:intro}

\textcolor{black}{The noise caused by snoring activity is an important problem in our society. The snoring noise can reach a sound level of $90$\,dB and have harmful implications, e.g., loss of productivity, attention deficit, and unsafe driving~\cite{chakravarthy2006application,chang2013active}. Recently, various studies have identified significant similarities between snoring and vocal signal~\cite{beck1995acoustic,pevernagie2010acoustics}. In fact, both of them present high-order harmonics preceded by a fundamental frequency in the spectrum~\cite{pevernagie2010acoustics}. The snoring activity is composed of two phases, i.e., inspiration and expiration. The power of the snoring signal is mostly concentrated on lower frequencies of the spectrum. In particular, the inspiration produces a signal between $100$\,Hz and $200$\,Hz, while the expiration is focused between $200$\,Hz and $300$\,Hz. Thus, the fundamental frequency, which must be deleted, is located between $100$\,Hz and $300$\,Hz.}

\textcolor{black}{In the literature, several approaches can be found for snoring attenuation. Passive solutions involve physical devices such as earplugs or special pillows~\cite{wei2010development} that may be troublesome for the user. Moreover, these techniques are ineffective at low frequencies and can be very expensive. In contrast, active noise control (ANC) systems can reduce low-frequency noises that passive approaches cannot attenuate. In particular, ANC techniques are based on the introduction of a secondary source that produces a signal capable of generating destructive interference in a desired area controlled by one or more microphones. ANC systems must be adaptive to follow the variations of the noise recorded at the error and reference microphones.
They are usually implemented using filtered-X least mean square (FxLMS)~\cite{morgan1980analysis} algorithm, where the estimate of the secondary path is used to calculate the output signal at the error microphone. Examples of FxLMS applications for active snoring cancellation can be found in~\cite{chakravarthy2006application,yenduri2006quiet,kuo2007real,cecchi2018real, nobili2021real,nobili2021efficient}.}

\textcolor{black}{However, snoring is a non-stationary signal that can cause issues during the adaptation process. Specifically, its irregular nature can result in signal absence, which in turn can negatively impact the performance of the adaptive algorithm. Therefore, to ensure active snoring cancellation, it is crucial to support it with a snoring activity detection algorithm that can identify the presence of snoring.}

\textcolor{black}{In the literature, deep learning algorithms for sound event detection and classification have also been applied to snoring audio signals. To this end, several studies have employed 2D convolutional neural networks (2D-CNNs) that rely on feature learning of time-frequency representations computed from fixed-length audio segments~\cite{khan2019deep,ansari2021deep,he2022novel}. In these studies, the high accuracy in snoring detection derives from both the acoustic features chosen and the wide signal analysis windows ($\geq1$\,s) that entail a slow decision response of the algorithm. This issue can be solved by sequential models that analyze the signal over short frames, such as 1D convolutional neural networks (1D-CNNs) and recurrent neural networks (RNNs). In~\cite{sun2019snorenet,hu2022auditory}, 1D-CNNs proved to be less performing than 2D-CNNs, but the low computational cost due to feature extraction from the raw audio signal makes them suitable for end-to-end systems. In~\cite{arsenali2018recurrent,lim2019classification}, RNNs exploited the features of past and present time-frequency representations of the audio signal over reduced temporal windows ($25-30$\,ms) for the snoring activity detection, confirming their effectiveness in sequential data analysis. Promising results have also been obtained from the combination of convolutional and sequential models, which together form convolutional recurrent neural networks (CRNNs). The studies described in~\cite{vesperini2020convolutional,jiang2020automatic} demonstrated that CRNNs with gated recurrent units (GRUs) or long short-term memory (LSTM) layers outperform 2D-CNNs in snoring detection. However, the performance of each approach is not easily comparable due to the different quantity, quality, and acquisition methods of the data used for training and testing the algorithms.}

\textcolor{black}{Given these premises, requirements such as reliability in signal classification and the capability to generalize in the presence of different background noises are some of the desired ones for an effective active snoring cancellation system.}
In this context, an enhanced system for the detection and active cancellation of snoring signals is presented. In particular, starting from the use of a CRNN for snoring activity detection, a delayless subband approach for active snoring cancellation has been improved, reporting good results in terms of convergence time and cancellation quality achieved.
\textcolor{black}{The paper is focused on the performance of the active snoring cancellation system with and without the aid of the snoring detection stage; therefore, since our interest is to evaluate the active snoring cancellation performance, the comparison of our snoring activity detection system with others in the literature is not addressed here because out of our scope, but it can be addressed in future work.}

\textcolor{black}{The paper is organized as follows. Section~\ref{sec:snoring} and Section~\ref{sec:anc} describe the definition of the algorithm for both snoring activity detection and active snoring cancellation, respectively. Experimental results are reported in Section~\ref{sec:exp}, where several results obtained with snoring signals are presented. Finally, conclusions are drawn in Section~\ref{sec:conc}.}

\section{Snoring Activity Detection}\label{sec:snoring}

\begin{figure}[t]
\centering
\includegraphics[width=0.82\columnwidth]{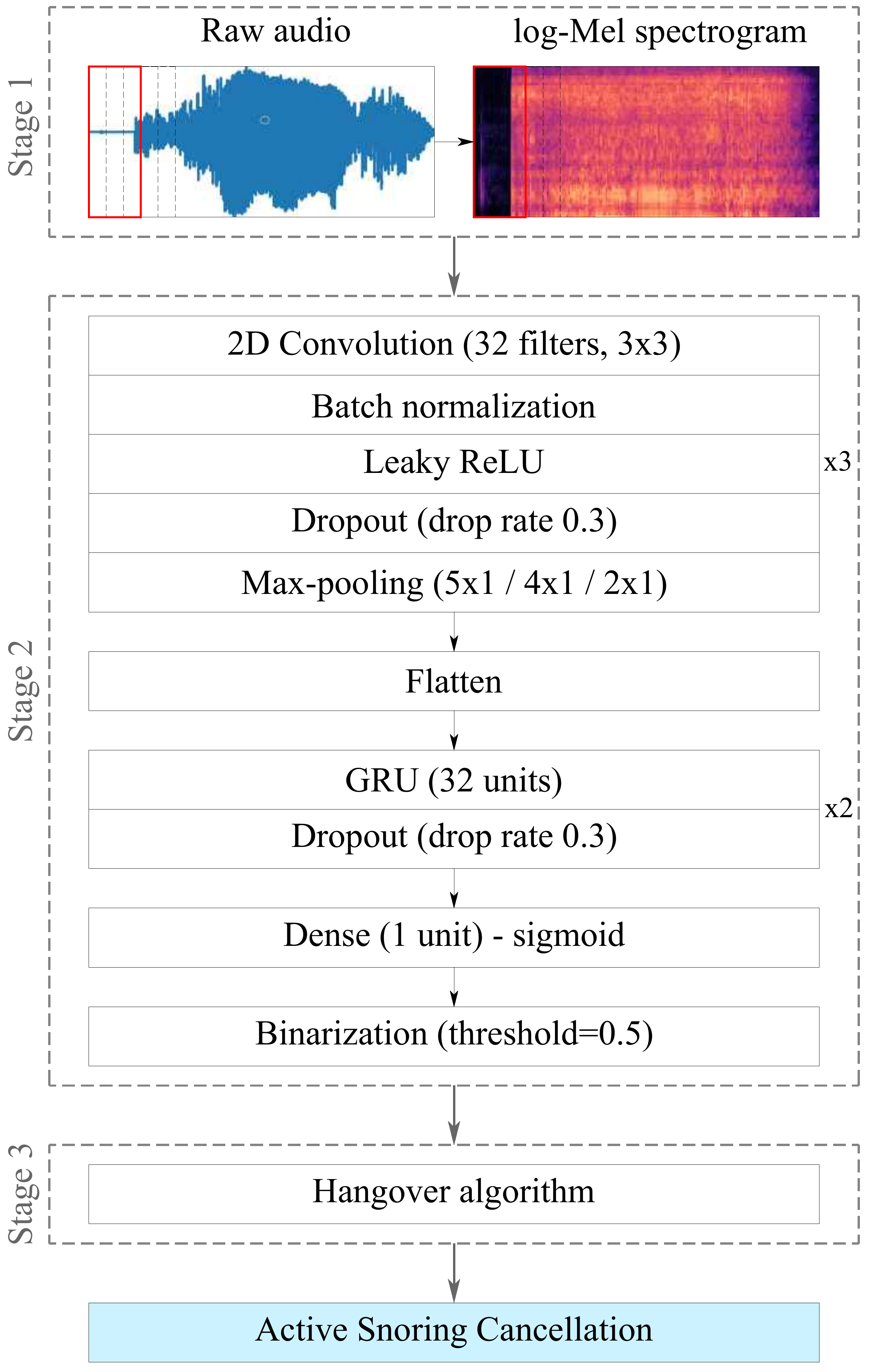}
\vskip -2mm
\caption{\textcolor{black}{Scheme of the SAD system.}}
\label{fig:sad}
\vskip -5mm
\end{figure}

\textcolor{black}{In this study, we address the snoring activity detection (SAD) methodology in three stages, as reported in Figure~\ref{fig:sad}. The first stage involves audio signal processing for acoustic feature computation. The second stage consists of data analysis using a CRNN for a binary snoring/non-snoring classification task, where a snoring event represents the positive class (label $1$), and all non-snoring events constitute the negative class (label $0$). Finally, in the third stage, the predictions produced by the neural network are post-processed with the \say{Hangover} algorithm.}
\textcolor{black}{This pipeline - binary classifier plus output filter (Hangover) - is common in Voice Activity Detection tasks.}

\textcolor{black}{More in detail, in the first stage, the stereo audio signal is turned into monophonic by channel averaging.}
Log-Mel spectrograms are computed, and $40$ log-Mel coefficients are extracted by using $30$\,ms-windows with a shift of $10$\,ms.

\textcolor{black}{The second stage involves the classification and is performed by the CRNN, which takes as input the log-Mel coefficients computed in the previous step.
The convolutional part of the CRNN comprises three consecutive blocks, each consisting of a convolutional layer, a batch normalization layer, a dropout layer, and a max pooling layer. In each block, convolutional layers have $32$ filters with size ($3$,$3$), and their output is normalized and regulated by the Leaky Rectified Linear Unit (Leaky ReLU)~\cite{xu2015empirical} activation function.
All dropout layers are characterized by a rate equal to $0.3$, while max-pooling layers have filters decreasing with each block, from ($5$,$1$) to ($4$,$1$) to ($2$,$1$).
The output is then flattened and passed to the recurrent part of the network, composed of two blocks. Each consists of a $32$-unit GRU layer with tanh and hard sigmoid activation functions to update and reset the gates, respectively, and a dropout layer with a drop rate of $0.3$.
Finally, a time-distributed feed-forward output layer, with a single neuron and sigmoid as activation function, returns predictions in the range [$0$,$1$], each one representing the probability that a frame is associated with a snoring event.
Then, the predictions are binary-encoded (\say{binarization}), using a threshold of $0.5$, so that they can be leveraged by the ASC algorithm.}

\textcolor{black}{In the third stage, the predictions output by the CRNN are post-processed with the Hangover algorithm presented in Algorithm~\ref{alg:hangover}.
It works with an input buffer, \textit{buffIn}, which acts as a FIFO (First-In First-Out) register that is automatically updated with a new sample every $10$\,ms, and takes as input the number of predictions in the input audio file $L$, the size of the input buffer $X$, and the number of predictions $k$ that we would like to use to characterize a snoring event.
When the input buffer is filled with the first $X$ samples, \textit{buffInFull()} returns the execution of the code to the caller; then the input buffer is read and a majority voting scheme is applied.
In particular, if the input buffer contains more zeros than ones, its content is copied into the output buffer, \textit{buffOut}. On the other hand, if it contains more ones than zeros, the Hangover algorithm considers the beginning of a snoring event by setting $k$ consecutive predictions to one. In this way, a snoring event is more likely to be characterized by all predictions equal to one. This method aims to decrease the number of sporadic false negatives (FNs) predictions (i.e., snoring occurrences erroneously classified as non-snoring) within a snoring sequence, which could degrade the ASC performance. Although this method is not robust against false positives (FPs), it is able to reduce FNs, which are those to which the ASC algorithm is most susceptible.}

  
        

        
        
                
                
                
            
            

\SetKwInput{KwInput}{Input}                
\SetKwInput{KwOutput}{Output}
\SetKw{KwBy}{by}
\SetKw{KwAnd}{and}
\begin{algorithm}[H]
\caption{Hangover algorithm}\label{alg:hangover}
\DontPrintSemicolon
  \KwInput{$L$, $k$, $X$}
  \KwOutput{$\mathbf{buffOut}$  \tcp{output buffer}}
  $outIdx \gets 0$ \tcp{index output buffer}
  
  $buffInFull(X)$ \tcp{returns when buffIn full}

 \While {$outIdx \le (L-1)$}{
    $\mathbf{buffIn} \gets readbuffIn()$

    $zeros \gets \textit{FindZeros}(\mathbf{buffIn})$ \tcp{Find n° of 0s}
        
    $ones \gets \textit{FindOnes}(\mathbf{buffIn})$ \tcp{Find n° of 1s}
    
    \If{$ones > zeros$}{ 

        $startIdx = outIdx$
    
        $\mathbf{buffOut}[startIdx - X : startIdx + 1] \gets \mathbf{I}(X,1)$

        
            
            

        $i \gets 1$
                
        \While {$1$}{
        
        $\_ \gets readbuffIn()$ \tcp{Discard reading}
        
            \If{$(i \le k - X)$ and $(outIdx \le L - 1)$}{

                $outIdx \gets startIdx + i$
            
                $\mathbf{buffOut}[outIdx] \gets 1$
                
                $i \gets i + 1$
            }
            \Else{
                $outIdx \gets outIdx + 1$
                
                $break$
            }
        }
    }
    \Else{
       \If{$outIdx = 0$}{
            $\mathbf{buffOut}[outIdx : outIdx + X] \gets \mathbf{buffIn}$
            
            $outIdx \gets outIdx + X$
        }
        \Else{
           $\mathbf{buffOut}[outIdx] \gets \mathbf{buffIn}[- 1]$

           $outIdx \gets outIdx + 1$
        }
    }
}
\end{algorithm}

\section{Active Snoring Cancellation}\label{sec:anc}
\begin{figure}
\centering
\includegraphics[width=\columnwidth]{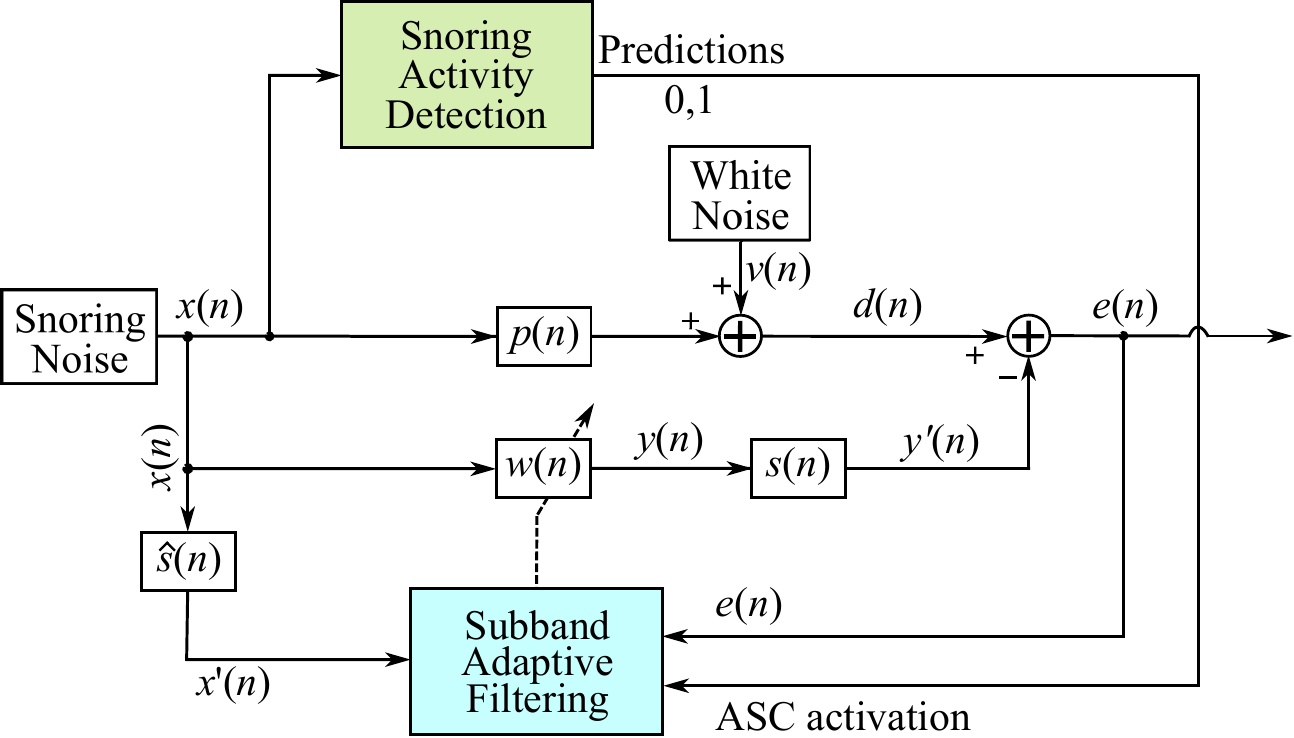}
\vskip -2mm
\caption{\textcolor{black}{Scheme of the ASC algorithm with SAD.}}
\label{fig:ANCscheme}
\vskip -5mm
\end{figure}

\textcolor{black}{Active Snoring Cancellation (ASC) is developed considering a feed-forward filtered-X configuration and a subband implementation as reported in \cite{nobili2021efficient}. Figure~\ref{fig:ANCscheme} shows the scheme of the algorithm. There is a reference microphone that picks up the
snoring source $x(n)$ and an error microphone that picks up the noise in the area to be quiet $e(n)$.  Then, a loudspeaker reproduces the interference signal $y(n)$ generated by $x(n)$ filtered with the adaptive filter $w(n)$ that represents the estimation of the primary path $p(n)$. The coefficients of this filter are produced by the subband adaptive filtering (SAF)
block on the basis of $x(n)$ filtered with the estimation of the path between the loudspeaker and the error microphone, i.e., the secondary path $s(n)$, the error $e(n)$, and snoring detection block predictions.}

\textcolor{black}{The SAF block has been developed considering a delayless subband adaptive filter algorithm as first proposed in~\cite{morgan1995delayless} and efficiently implemented in~\cite{nobili2021efficient,nobili2021real}. In particular, the signal $x'(n)$ and the error $e(n)$ are decomposed in subband by an analysis filter-bank, as $x'_k(n)$ and $e_k(n)$ for each $k$-th subband, respectively. The weights of the $k$-th subband $\textbf{w}_k^{SAF}(n)$ are updated following the normalized least mean square (NLMS) algorithm as
\begin{equation}
    \textbf{w}_k^{SAF}(n+1) = \textbf{w}_k^{SAF}(n) + \mu_w \frac{\textbf{x}'^{\ast}_k(n)e_k(n)}{\alpha + ||\textbf{x}'_k(n)||^2 },
\end{equation}
where $\textbf{x}'^{\ast}_k(n)$ is the complex conjugate of the input signal of the $k$-th subband $x'_k(n)$, $\mu_w$ is the step size, and $\alpha$ is a small coefficient that avoids division by zero. The fullband filter $w(n)$ of length $N$ is obtained by stacking all the subband weights following the steps below:
\begin{itemize}
\item the subband weights are reported in the frequency domain by $(N/D)$-point fast Fourier transform (FFT), with $D=M/2$ the decimation factor and $M$ the number of subbands;
\item the first half of the array representing the fullband filter is calculated by stacking the complex samples of FFT;
\item the rest of the array is obtained by the complex conjugate reversed version of the first half and the central point is set to zero. 
\item the fullband filter is computed by a $N$-point inverse FFT of the array. 
\end{itemize}}
\textcolor{black}{The SAF algorithm is activated when the SAD algorithm provides a prediction of snore presence.}

\section{Experimental Results}\label{sec:exp}
\subsection{Dataset}\label{sec:dataset}

\textcolor{black}{The A3-Snore dataset~\cite{vesperini2020convolutional} has been selected for the experimental phase. It is a collection of audio files containing snoring events emitted by two male volunteers aged 48 and 55 during overnight sleep. The recording setup is a ZOOM-H1 Handy Recorder with two unidirectional microphones oriented perpendicularly. Acquisitions were made in a single room measuring $4 \times 2.5$ m with the sensors positioned near the snorer's head. The corpus includes almost $7$\,h of audio material split into $10$-minute segments, selected according to the highest frequency of snoring events associated with each volunteer (\say{snorer 1} and \say{snorer 2}). All audio files, characterized by wav format, are stereophonic with a sampling rate of $44.1$\,kHz and $16$-bit encoding. A metadata file reports annotations of the start and end timestamps of snoring events with a resolution of $1$ second. The dataset is organized into two folders, each associated with a snorer, with an unbalanced distribution between snoring and non-snoring events. Table~\ref{tbl_dataset} summarizes the composition of the A3-Snore audio collection.} \textcolor{black}{Files associated with Snorer 1 have been used for the training set, whereas Snorer 2's files have been split with a ratio of 50\% and used for validation and test sets.}

\begin{table}[b!]
\vskip -5mm
\caption{\textcolor{black}{Statistics of the A3-Snore dataset.}}\label{tbl_dataset}
\vspace{0.1 cm}
\centering
\resizebox{.9\columnwidth}{!}{  
\begin{tabular}{|c|c|c|c|c|}
\hline
Snorer & \begin{tabular}[c]{@{}c@{}}Number \\ of files\end{tabular} & \begin{tabular}[c]{@{}c@{}}Total \\ duration\\ {[}s{]}\end{tabular} & \begin{tabular}[c]{@{}c@{}}Snoring \\ duration\\ {[}s{]}\end{tabular} & \begin{tabular}[c]{@{}c@{}}Snoring \\ ratio\\ {[}\%{]}\end{tabular}      \\ \hline
1        & 18       & 10\,800       & 1127        & 10.4      \\ 
2        & 23       & 13\,800       & 2017        & 14.6     \\ 
Total    & 41       & 24\,600       & 3144        & 12.8     \\ \hline
\end{tabular}
}
\end{table}

\subsection{Snoring Activity Detection}

\textcolor{black}{In the experiments, training was performed in a supervised manner for $500$ epochs \textcolor{black}{by monitoring the Average Precision (AP) - also known as the area under the precision-recall curve (AUC-PR) - on the validation set, and exploiting the early-stopping strategy} to arrest the learning process when the model does not improve for $20$ consecutive epochs.}

\textcolor{black}{An adaptive learning rate according to the AdaDelta~\cite{zeiler2012adadelta} optimization algorithm was selected, with an initial value equal to $1$ and a decay rate of $0.95$. The binary cross-entropy was used as the loss function. The experiments were carried out on an NVIDIA DGX Station A100 with dual 64-Core AMD EPYC 7742 @3.4 GHz and eight NVIDIA A100-SXM4-40 GB GPUs. The server was running Ubuntu 20.04.3 LTS. The neural network has been implemented with the Tensorflow~\cite{abadi2016tensorflow} deep learning framework.}

\textcolor{black}{CRNN classification performance was evaluated considering the AP, obtaining a value equal to $77.54$\%.}

\textcolor{black}{For what concerns the Hangover algorithm, the size $X$ of the input buffer has been chosen in order to reduce the number of FNs while keeping the latency as low as possible. Moreover, since the Hangover algorithm applies a majority voting scheme, $X$ should be an odd number.
We found the right trade-off by setting $X=3$; in this way, the post-processing algorithm is able to improve the CRNN output while maintaining a relatively low latency (i.e., 30\,ms).
Since for a 10-minute audio file we have $60\,001$ predictions, we set $L=60\,001$, whereas $k$ has been set equal to $100$.}

\begin{figure}[t]
\centering
\subfigure[\label{fig:post}]{\includegraphics[width=0.85\columnwidth]{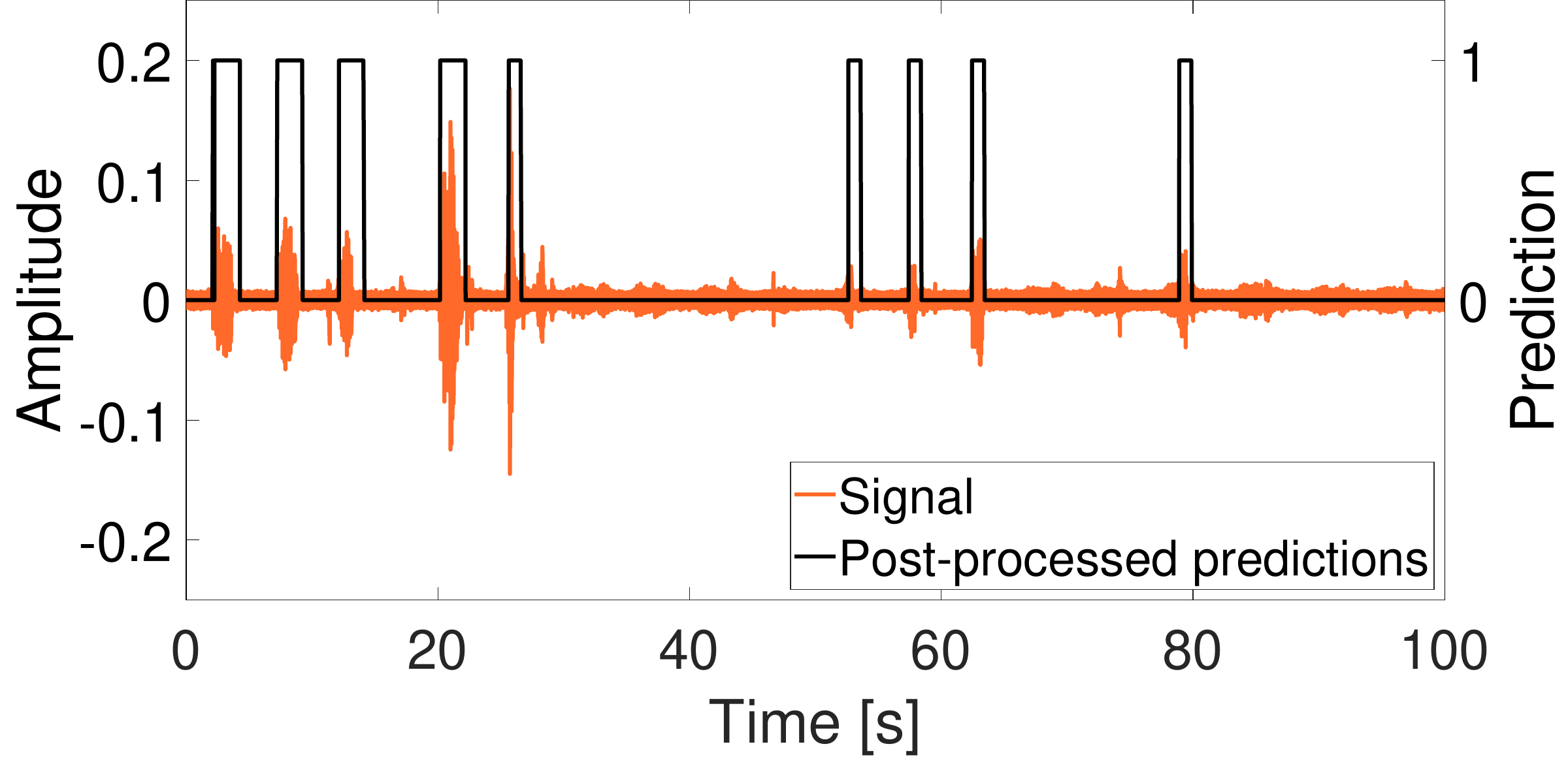}}
\subfigure[\label{fig:raw}]{\includegraphics[width=0.85\columnwidth]{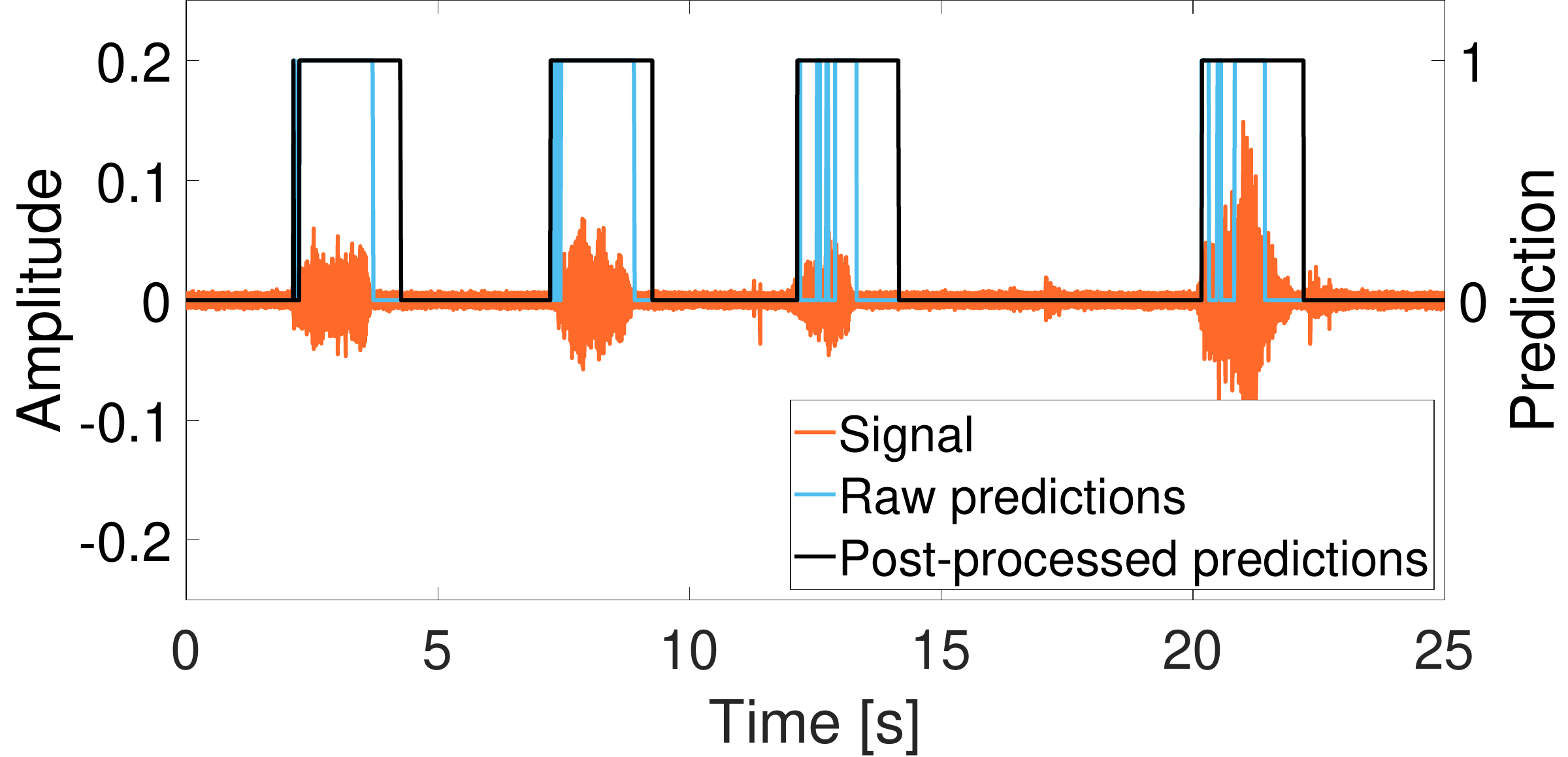}}
\vskip -2mm
\caption{
\textcolor{black}{Predictions post-processed by the Hangover algorithm of an audio signal~\subref{fig:post}, and their difference with respect to raw predictions before the processing stage~\subref{fig:raw}.}
}
\label{fig:signal_pred}
\vskip -5mm
\end{figure}

In order to evaluate the performance of the overall active snoring detection system also from a graphical perspective, we report in Fig.~\ref{fig:signal_pred}\subref{fig:post} a 100-second excerpt of an audio signal employed in testing and the associated predictions generated by the overall snoring activity detection system, after the post-processing stage.
Moreover, in order to also visualize the Hangover Algorithm performance, Fig.~\ref{fig:signal_pred}\subref{fig:raw} shows the binary predictions output of the CRNN before and after the post-processing stage; the time interval is more limited to better highlight the difference.

\subsection{Active Cancellation with Snoring Activity Detection}
\textcolor{black}{The presented ASC algorithm has already been validated in \cite{nobili2021efficient,nobili2021real}, by comparing its performance with the state-of-the-art algorithm of \cite{zhang2001cross}, considered as reference. In this paper, the ASC algorithm is improved by applying the SAD, and the experiments are mainly focused on evaluating the performance of the system with and without SAD.} 
Starting from the snoring signals of the dataset described in Section~\ref{sec:dataset}, primary path $p(n)$ and secondary path $s(n)$ are simulated considering responses measured in a semi-anechoic chamber from the setup of~\cite{cecchi2018real}. Since $p(n)$ and $s(n)$ are modeled as FIR filters with a length of $256$ samples, the length of the adaptive filter $w(n)$ is set to $512$ taps. For the subband structure, the length of the prototype filter is $256$ samples, the number of subbands is $M=64$, and the step size is $\mu=0.03$.
The performance of the proposed system has been evaluated in terms of primary path estimation, varying the signal-to-noise ratio (SNR) of the signal $d(n)$ (cf. Figure~\ref{fig:ANCscheme}).

\begin{figure}[t!]
\centering
\subfigure[\label{fig:IRest}]{\includegraphics[width=0.8\columnwidth]{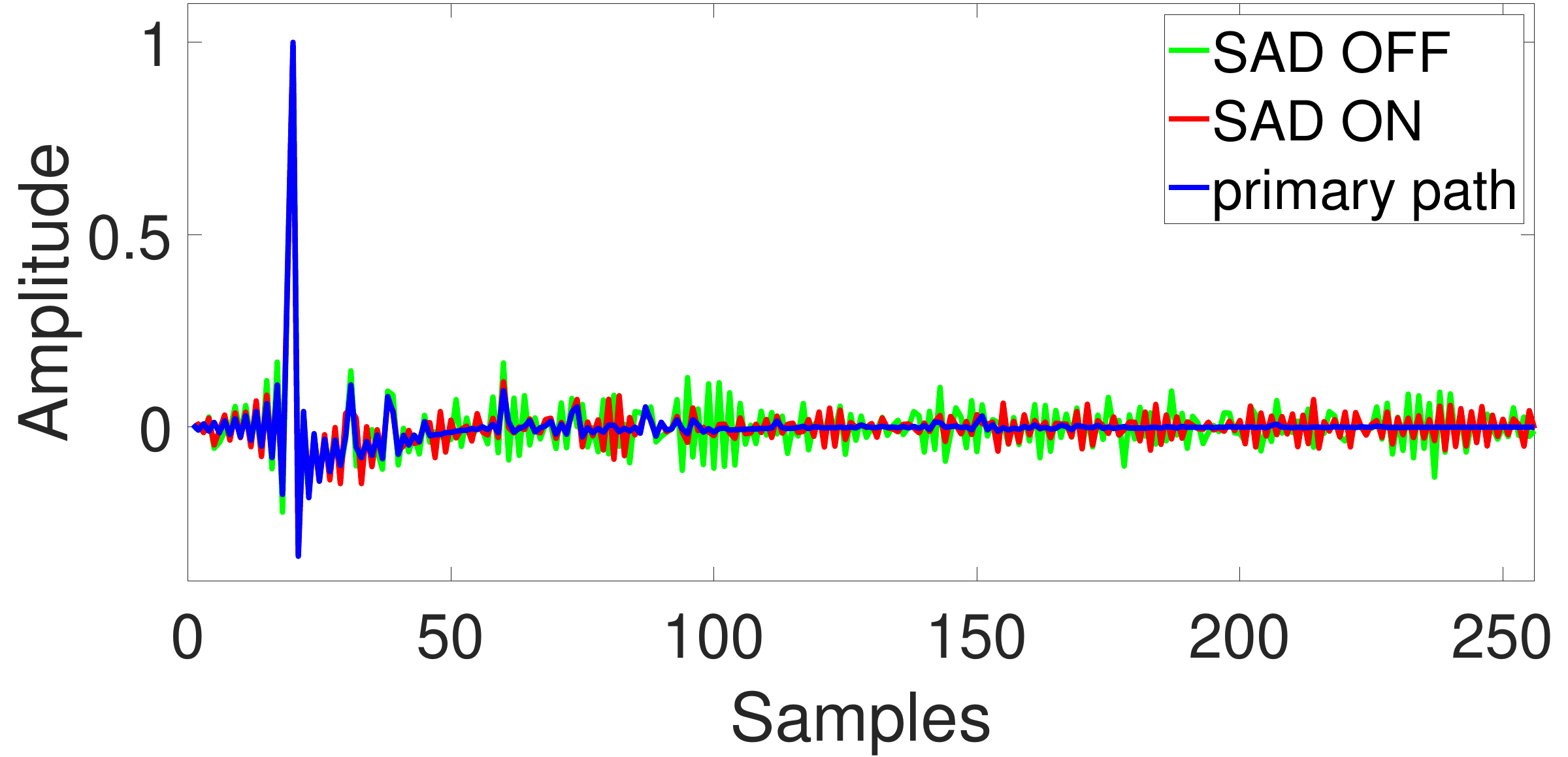}}
\subfigure[\label{fig:FRest}]{\includegraphics[width=0.8\columnwidth]{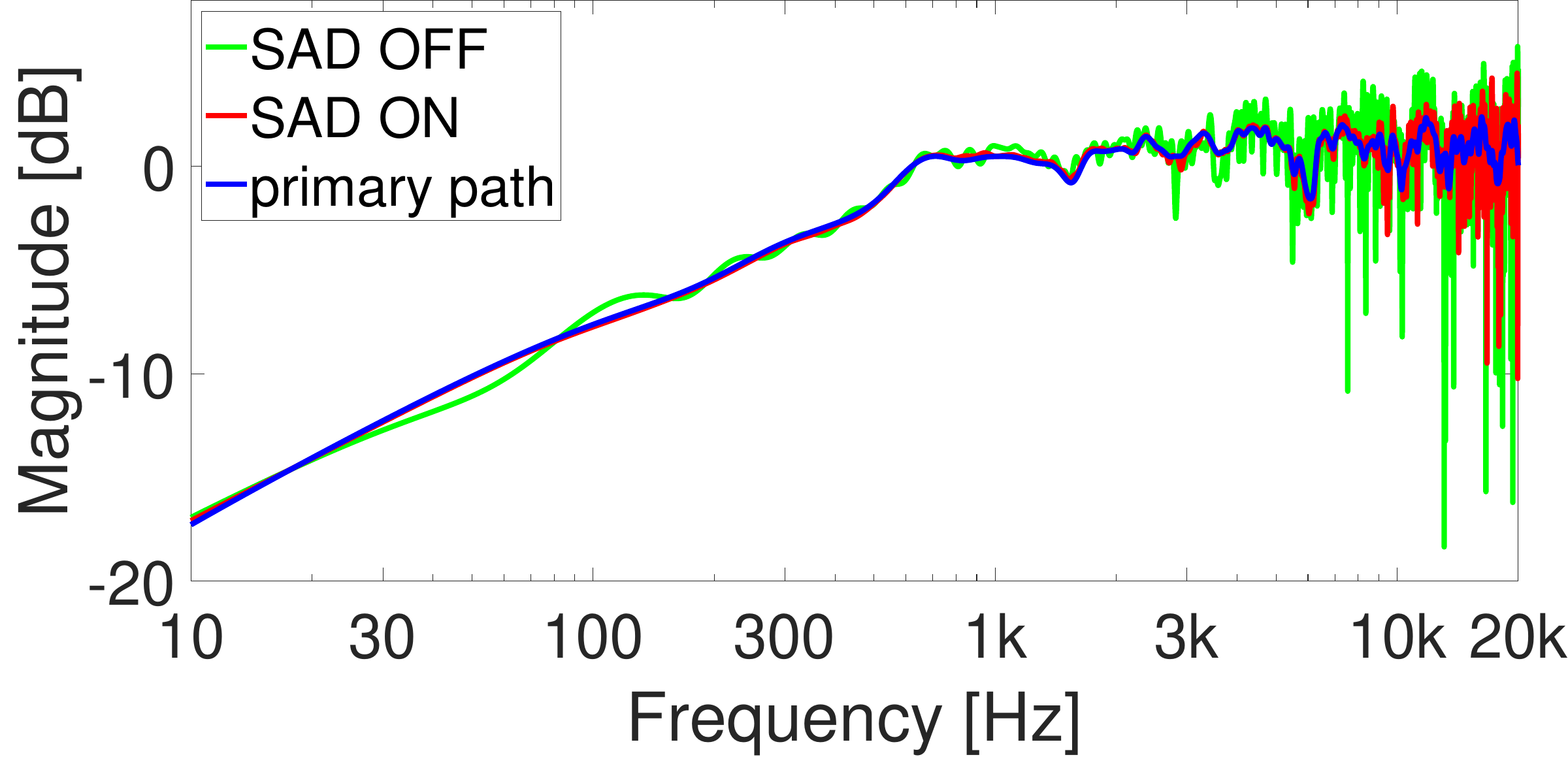}}
\vskip -2mm
\caption{Comparison between the measured primary path, the primary path estimated in the case of SAD OFF and SAD ON, \subref{fig:IRest} in the time domain and \subref{fig:FRest} in the frequency domain, considering an input signal with $\text{SNR}=10$\,dB.}
\label{fig:estimation}
\vskip -5mm
\end{figure}

The primary path estimated by the ASC with the SAD is compared with the one estimated without SAD and with the measured primary path. Figure~\ref{fig:estimation} shows the obtained results considering $\text{SNR}=10$\,dB. The difference between the estimated responses and the measured one is evaluated by the log-spectral distance (LSD), in the frequency domain, and by the misalignment, in the time domain. \textcolor{black}{The LSD evaluates the spectral difference between two frequency responses 
\cite{cecchi2022using}. Similarly, the misalignment evaluates the difference between the measured and the estimated path in the time domain and gives a measure of the convergence rate \cite{nobili2021real,nobili2021efficient}.}
Denoting the measured primary path as $p(n)$, the estimated primary path as $w(n)$, and their respective transfer functions as $P(k)$ and $W(k)$, the LSD is computed as
\begin{equation}\label{eq:lsd}
    \text{LSD}=\sqrt{\frac{1}{k_2-k_1+1}\sum_{k=k_1}^{k_2}\Bigg[10\log_{10}\frac{\big|P(k)\big|^2}{\big|W(k)\big|^2}\Bigg]^2},
\end{equation}
where $k_1$ and $k_2$ delimit the frequency range within which the LSD is estimated, defined as $B=[k_1\frac{f_\text{s}}{K},\, k_2\frac{f_\text{s}}{K}]=[100\,\text{Hz},\,20\,\text{kHz}]$, with $K=4096$ the number of frequency bins for the FFT computation, and $f_\text{s}=44.1$\,kHz the sampling frequency.
The misalignment is calculated as
\begin{equation} \text{MIS}=20\log_{10}\frac{||p(n)-w(n)||}{||p(n)||}.
\end{equation}
Table~\ref{tab:lsdmis} shows the values of the LSD and the misalignment considering signals with different SNR levels. The estimation performance improves with the SNR increase both with and without SAD and in terms of both LSD and misalignment. The lowest values of the LSD are obtained when the SAD is applied, i.e., when the adaptation algorithm of the ASC is executed only when the snoring signal is detected by the SAD. This result is confirmed by Figure \ref{fig:FRest}, where the magnitude frequency response of the primary path is well estimated up to 10\,kHz with SAD, while the frequency response estimated without SAD deviates from the measured one for all the frequency spectrum. Differently, the difference in the misalignment of the two cases is more difficult to recognize. In fact, looking at Figure \ref{fig:IRest}, the main peak of the impulse response is rightly detected both with and without SAD, but both cases introduce some late reflections not present in the measured impulse response.  

\begin{table}[t!]
\caption{\textcolor{black}{Values of the LSD and the misalignment obtained considering SAD OFF and SAD ON for different SNR values of the input signal. The LSD is calculated in the frequency range of [$100$\,Hz--$20$\,kHz].}}\label{tab:lsdmis}
\vspace{0.2 cm}
\centering
\resizebox{0.9\columnwidth}{!}{  
\begin{tabular}{|c|c|c|c|c|}
\hline
\multirow{2}{*}{SNR [dB]}  &\multicolumn{2}{c|}{LSD [dB]} &\multicolumn{2}{c|}{Misalignment [dB]}\\
\cline{2-5}
& SAD OFF & SAD ON & SAD OFF & SAD ON\\
\hline
10 & 0.79 & \textcolor{black}{\textbf{0.72}} & -4.05 & \textcolor{black}{\textbf{-6.05}}\\
15 & 0.49 & \textcolor{black}{\textbf{0.37}} & -10.11 & \textcolor{black}{\textbf{-12.51}}\\
20 & 0.25 & \textcolor{black}{\textbf{0.21}} & \textbf{-16.20} & \textcolor{black}{-14.86}\\
\hline
\end{tabular}
}
\end{table}

\section{Conclusions}\label{sec:conc}
\textcolor{black}{In this paper, an enhanced system that combines detection and active cancellation of snoring signals has been proposed. For snoring activity detection, a convolutional recurrent neural network fed by log-Mel coefficients has been implemented to classify snoring and non-snoring events. For active snoring cancellation, a feed-forward filtered-X configuration based on a delayless subband adaptive filter algorithm has been developed. The combined use of the two algorithms results in a single improved system for ASC. This work is a preliminary study that offers large room for improvement. For the SAD, more performing neural architectures based on unsupervised or semi-supervised deep learning strategies coupled with larger and more challenging datasets can be explored.}
\textcolor{black}{The ASC can be improved by introducing non-uniform subband structures and different environments with different reverberations could be taken into account to test the proposed system.}


\setstretch{0.93}
\bibliographystyle{IEEEbib}
\bibliography{refs}

\end{document}